\documentclass[reprint,aps,pra,showpacs,
superscriptaddress,longbibliography]{revtex4-1}
\usepackage{graphicx}
\usepackage{amsmath}
\usepackage{amssymb}
\usepackage{epstopdf}
\usepackage{amsfonts}
\usepackage{dcolumn}
\usepackage{bm}
\usepackage{color}

\begin{document}

\title{Dynamics of small trapped one-dimensional Fermi gas under
oscillating magnetic fields}
\author{X. Y. Yin}
\affiliation{Department of Physics, The Ohio State University, Columbus,
Ohio 43210, USA}
\author{Yangqian Yan}
\affiliation{Department of Physics and Astronomy, Washington State University, Pullman,
Washington 99164-2814, USA}
\affiliation{Department of Physics, Indiana University Purdue University Indianapolis (IUPUI),
Indianapolis, Indiana 46202, USA}
\author{D. Hudson Smith}
\affiliation{Department of Physics, The Ohio State University, Columbus,
Ohio 43210, USA}

\date{\today }

\begin{abstract}
Deterministic preparation of an ultracold harmonically trapped one-dimensional
Fermi gas consisting of a few fermions has been realized by the Heidelberg group.
Using Floquet formalism, we study the time dynamics of two- and three-fermion 
systems in a harmonic trap under an oscillating magnetic field. 
The oscillating magnetic field produces a time-dependent interaction strength
through a Feshbach resonance.
We explore the dependence of these dynamics on the frequency of the oscillating magnetic
field for non-interacting, weakly interacting, and strongly interacting systems. 
We identify the regimes where the system can be
described by an effective two-state model and an effective three-state model.
We find an unbounded coupling to all excited states at the infinitely strong interaction limit
and several simple relations that characterize the dynamics.
Based on our findings, we propose a technique for driving transition 
from the ground state to the excited states using an oscillating magnetic field. 

\end{abstract}

\pacs{03.75.Ss, 05.30.Fk}
\maketitle

\section{Introduction}
\label{sec_intro}

Recent advances in ultracold atom experiments have allowed for the
deterministic preparation of few-atom systems
in a one-dimensional harmonic trap~\cite{jochim11, jochim13}. 
Feshbach resonance and
confinement induced resonance (CIR) provide a convenient way to tune
the effective interatomic interaction strength~\cite{tiesinga_rmp, olshanii98}.
The few-body energy spectra and eigenstates of fermions in a 1D trap
have been explored by recent theoretical 
works~\cite{ebrahim13, harshman12, lewenstein13, zinner14, parish15}.
These cold atom systems provide a clean
and controlled platform for studying the tunneling dynamics
and pairing of a few atoms~\cite{jochim13b, jochim15}.
Many fundamental quantum mechanics problems of great theoretical interest
can nowadays be directly prepared in such a platform. 
Among the most important of these problems are driven quantum systems,
which are not only of theoretical interest but also have application in 
quantum chemistry~\cite{chapter5, floquet_njp, dalibard14, floquet_review}.
Time-dependent external driving introduces additional energy scales,
which are associated with the driving frequency and driving strength, to a
quantum system and can therefore generate new phenomena. 
It is especially interesting to see the interplay between the
energy scales associated with
the external harmonic confinement, the interparticle interactions, and the 
oscillating magnetic field.

In the experiments of the Heidelberg group, 
systems consisting of 1 to 10 lithium atoms 
are prepared in a
highly elongated optical dipole trap~\cite{jochim11, jochim13}. 
The impurity is a single lithium atom which occupies a different
hyperfine state than the majority lithium atoms.
Such systems can be prepared with high fidelity in the molecular branch when the 
coupling constant is negative and in the
the upper branch when the
coupling constant is positive~\cite{jochim11, jochim13, blume15}. 
Present studies of the interaction energies and tunneling dynamics
are mostly based on the ground state of the 
systems~\cite{jochim11, jochim13, jochim13b}. 
The access to excited states will allow for a higher degree of tunability.
Tunneling dynamics may be studied when the system
is initially prepared in an excited state. Also, the degenerate manifolds in the
excited states may be accessed, and the coupling within the manifold
could be studied.
The oscillation of a magnetic field near a Feshbach resonance has been used to associate
atoms into dimers and to dissociate dimers into atoms
by various experimental groups~\cite{wieman05, jin07}. 
Such transitions have been investigated in various theoretical 
works~\cite{borca03, klaus06, plata15, braaten15}.
These works motivate us to propose a technique for populating the excited states in
trapped one-dimensional, two-component Fermi gas using an oscillating magnetic
field.

In this work,
we study the dynamics of two- and three-fermion systems
with time-dependent interaction strength generated by 
an oscillating magnetic field in the non-interacting limit, the weakly-interacting
regime, and the infinitely strong interaction limit. 
We focus on the weak-driving limit, where the energy
scale associated with the driving strength is much smaller than 
the energy scale associated with the harmonic trap.
We also focus on the regime where the
driving frequency is comparable to
the trapping frequency. 
Systems subject to strong, high-frequency driving exhibit vastly different
behavior from the dynamics discussed in this work, including
energy cascades and quantum turbulence~\cite{ho16}.
We will show that the time dynamics depend crucially on whether the driving
frequency is on resonance with a transition between two eigenstates. 
We will also demonstrate a distinct difference
between the dynamics for the non-interacting system and the 
infinitely strongly interacting system.
Specifically, we find an unbounded coupling to all excited states in the
infinitely strongly interacting system.
In this case, we find the time and magnitude each excited state is populated.
We also identify that, for certain parameter combinations,
the system can be accurately described by a two-state or three-state model. 
These findings naturally lead to a strategy for optimally
driving transitions between the ground state and the excited states.

The remainder of this paper is organized as follows. Section~\ref{sec_theory}
discusses the theoretical framework,
including the system Hamiltonian and the Floquet formalism. The 
relation between the parameters
in the model Hamiltonian and the experimental works is also discussed.
Section~\ref{sec_result} presents our results for the dynamics of two- and three-
fermion systems in the non-interacting limit, weakly interacting regime, and the
the infinitely strong interaction limit.
Section~\ref{sec_conclusion} concludes.

\section{System Hamiltonian and general consideration}
\label{sec_theory}

We consider a single impurity with one or two
identical fermions in a one dimensional
harmonic trap, denoted as (1,1) and (2,1) systems,
respectively. 
The impurity interacts with the fermions
through a zero-range two-body potential with coupling constant $g$,
\begin{eqnarray}
V_\text{2b}(z_{j0})=g\delta(z_j-z_0),
\end{eqnarray}
where $z_0$ is the position of the impurity and
$z_j$ the position of the identical fermions ($j=1$ or $2$).
As we will show, the effect of the oscillating magnetic field is
contained in the time dependence of the coupling constant.
For the ($N$,1) system confined in a harmonic trap with angular trapping
frequency $\omega_z$, the Hamiltonian reads
\begin{eqnarray}
\label{eq_hamiltonian}
H=\sum_{j=1}^{N} H_\text{ho}(z_j)+H_\text{ho}(z_0)+
\sum_{j=1}^{N}V_\text{2b}(z_{j0}).
\end{eqnarray}
The single particle harmonic oscillator Hamiltonian $H_\text{ho}(z)$ is given by
\begin{eqnarray}
H_\text{ho}(z)=-\frac{\hbar^2}{2m}\frac{\partial^2}{\partial z^2}
+\frac{1}{2}m\omega_z^2 z^2.
\end{eqnarray}

In cold atom experiments, the effective 1D trap is often created through
a cigar shaped trapping potential with the radial trapping frequency $\omega_\rho$
much greater than the axial trapping frequency $\omega_z$.
Near a confinement induced resonance (CIR), the 1D coupling constant $g$ is related to the 3D 
scattering length $a_\text{3d}$
by~\cite{olshanii98}
\begin{eqnarray}
\label{eq_g1d}
g=\frac{2\hbar^2 a_\text{3d}}{m a_\rho^2} 
\frac{1}{1-|\zeta(1/2)| a_\text{3d}/(\sqrt{2}a_\rho)},
\end{eqnarray}
where $a_\rho=\sqrt{\hbar/(m \omega_\rho)}$ is the harmonic oscillator 
length of the tight
confining direction, $m$ is the atomic mass, $a_\text{3d}$ is the 3D scattering
length, and $\zeta(x)$ denotes the Riemann-Zeta function.
Near a Feshbach resonance, the 3D scattering length $a_\text{3d}$ is related to the
magnetic field $B$ by~\cite{feshbachli}
\begin{eqnarray}
\label{eq_a3d}
a_\text{3d}(B)=a_\text{bg}\left( 1-\frac{\Delta}{B-B_0} \right),
\end{eqnarray}
where $B_0$ is the Feshbach resonance position,
$\Delta$ is the
width of the resonance, and $a_\text{bg}$ is the background scattering length.
The 1D coupling constant $g$ diverges at the value of the magnetic field
\begin{eqnarray}
B_\text{CIR}=B_0+\Delta-\frac{\Delta}{1-\frac{a_\text{bg}|\zeta(1/2)|}{\sqrt{2}a_\rho}}.
\end{eqnarray}

We consider magnetic field $B(t)$ oscillating around a constant $\bar{B}$,
\begin{eqnarray}
\label{eq_Bt}
B(t)=\bar{B}+b\cos(\omega t),
\end{eqnarray}
where $\omega$ is the oscillating frequency.
We remark that in the weak-driving limit considered in this work,
the oscillating magnetic field with form $B(t)=\bar{B}+b\sin(\omega t)$
will lead to almost identical dynamics to the case considered in this work.
The reason is that in the weakly-driving limit, the time scale defined by
driving strength is much longer than the driving period $T=2\pi/\omega$.
Hence, the micro-motion, shown as small oscillation with frequency $\omega$,
is negligible compared to the dynamics studied in this work.
The separation of the dynamics on different time scales has been discussed
in Ref.~\cite{floquet_review}. In terms of their language, 
we focus on the non-stroboscopic dynamics.

In the non-interacting limit and the weakly-interacting regime,
the constant magnetic field $\bar{B}$ is far enough away from $B_\text{CIR}$
such that the coupling constant $g$ satisfies the condition
$g\ll a_z E_\text{ho}$, where $a_z=\sqrt{\hbar/(m\omega_z)}$ and
$E_\text{ho}=\hbar\omega_z$ are
the harmonic oscillator length and harmonic oscillator energy for the 1D trap,
respectively.
Combining Eqs.~(\ref{eq_g1d}), (\ref{eq_a3d}), and (\ref{eq_Bt}) and expanding
the resulting time-dependent 1D coupling constant $g(t)$ in terms of $b$
to the first order, we find
\begin{eqnarray}
\label{eq_gt}
g(t)\approx\bar{g}+d \cos(\omega t),
\end{eqnarray}
where the average coupling constant $\bar{g}$ is given by replacing 
$a_\text{3d}$ in Eq.~(\ref{eq_g1d})  with  $a_\text{3d}(\bar{B})$
given by Eq.~(\ref{eq_a3d}), and the driving strength $d$ is given by
\begin{eqnarray}
d=\frac{8\hbar^2 a_\text{bg}\Delta b}{m[2a_\rho(\bar{B}-B_0)-
\sqrt{2}|\zeta(1/2)| a_\text{bg}(\bar{B}-B_0+\Delta)]^2}.
\end{eqnarray}
In the $\bar{g}\rightarrow\infty$ limit, $\bar{B}$ is fixed at $B_\text{CIR}$.
Expanding the inverse 1D coupling constant $1/g(t)$ around $0$ and keeping terms up to
first order in $b$, we find
\begin{eqnarray}
\label{eq_invgt}
\frac{1}{g(t)}\approx h \cos(\omega t),
\end{eqnarray}
where $h$ is given by
\begin{eqnarray}
h=-\frac{m(\sqrt{2}a_\rho-a_\text{bg} |\zeta(1/2)|)^2b}{4\hbar^2 a_\text{bg} \Delta}.
\end{eqnarray}

\begin{figure}[htbp]
\includegraphics[angle=0,width=80mm]{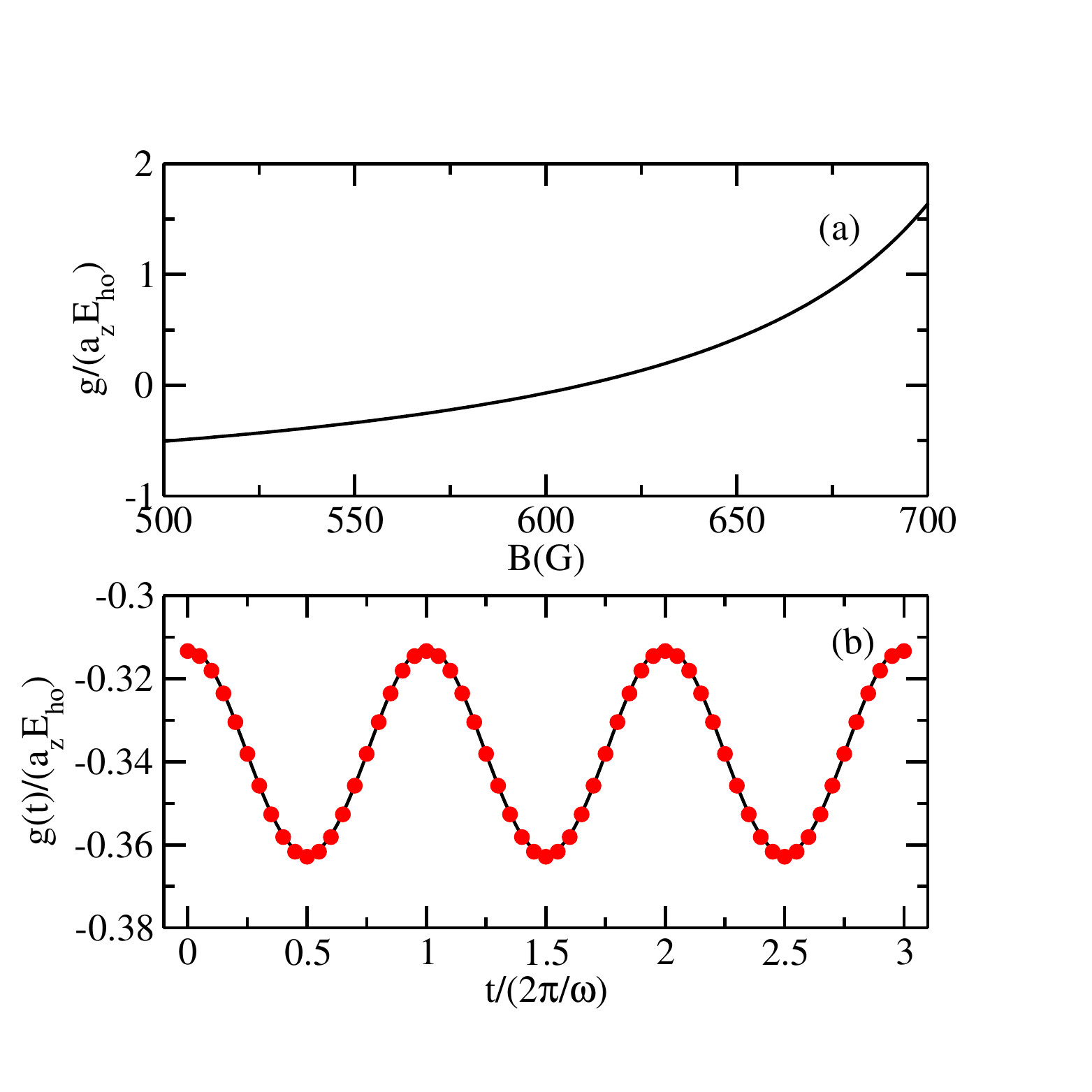}
\caption{(Color online) Panel (a) shows the 1D coupling constant $g$
as a function of the magnetic field $B$, 
calculated using Eqs.~(\ref{eq_g1d}) and (\ref{eq_a3d}).
Panel (b) shows $g(t)$ as a function of time for an 
oscillating magnetic field $B(t)=\bar{B}+b\cos(\omega t)$
with $\bar{B}=550\,$G and $b=6\,$G. 
The solid line is calculated from Eqs.~(\ref{eq_g1d})-(\ref{eq_Bt}).
The dots are calculated from the first order expansion given in Eq.~(\ref{eq_gt}).}
\label{fig_g1d}
\end{figure}

To test the validity of the approximations in Eq.~(\ref{eq_gt}) and (\ref{eq_invgt})
in the experimentally relevant regimes,
we consider two hyperfine states of ${}^6$Li,
referred to as $|2\rangle=|F=1/2,m_F=-1/2\rangle$ and
$|3\rangle=|F=3/2,m_F=-3/2\rangle$ as an example.
We consider the trapping frequencies implemented in Ref.~\cite{jochim13}, i.e.
$\omega_\rho=2\pi \times 14.22(0.35)$ kHz and $a_z=2\pi \times 1.488 (0.014)$ kHz.
The solid line in Fig.~\ref{fig_g1d}(a) shows the 1D coupling constant $g$
as a function of the magnetic field $B$, 
calculating using Eq.~(\ref{eq_g1d}) and (\ref{eq_a3d})
with $a_\text{bg}$, $B_0$, and $\Delta$ adapted from Ref.~\cite{feshbachli}.
We note that the solid line differs slightly from the the 
experimentally determined grey curve shown
in Figure S3 of Ref.~\cite{jochim13}. This is partly due to the trap calibration
and other nearby resonances. 
A detailed analysis of trap calibration is provided in Ref.~\cite{ebrahim15}.
The small difference between the solid line in Fig.~\ref{fig_g1d}(a)
and the one measured from the experiments does not affect the analysis of the 
validity regime of approximations in Eq.~(\ref{eq_gt}) and (\ref{eq_invgt}). 
A quantitative comparison
of the results of this paper and future experiments will require experimental
determination of $\bar{g}$ and $d$ from $\bar{B}$ and $b$
instead of using  Eqs.~(\ref{eq_g1d}) and (\ref{eq_a3d}) directly.

Figure~\ref{fig_g1d}(b) shows the time-dependent coupling constant $g(t)$
resulting from the time-dependent magnetic field in Eq.~(\ref{eq_Bt})
with $\bar{B}=550\,$G and $b=6\,$G. For this choice of parameters,
the first order approximation to $g(t)$ in Eq.~(\ref{eq_gt}) is accurate 
to within 3\%.
The comparison shows that the first order expansion provides an excellent description
of the time dependence of $g(t)$ in the weakly-interacting regime.
Equation (\ref{eq_gt}) still works reasonably well
even if the interaction energy is comparable to the scale defined by the trap,
i.e, $g(t)$ is comparable to $a_z E_\text{ho}$.
For oscillating magnetic field with $\bar{B}=680G$ and $b=4G$, 
$g(t)$ is averaged at $0.99a_z E_\text{ho}$
and oscillates between $1.09a_z E_\text{ho}$ and 
$0.89a_z E_\text{ho}$. In this case, the difference between the 
first order expansion and the exact result is
less than $5\%$ of the oscillating magnitude $d$.
Similarly, we examine the validity regime of the expansion in the $\bar{g}\rightarrow\infty$ limit by 
comparing $1/g(t)$ obtained from
the first order expansion given in Eq.~(\ref{eq_invgt}) and the exact result
calculated from Eq.~(\ref{eq_g1d})-(\ref{eq_Bt}). We find that 
the first order expansion
differs from the exact result by less than $1\%$ for $b<1.5G$.

We solve the time-dependent problem using the standard
Floquet formalism~\cite{floquet_original, floquet_njp}. We rearrange the Hamiltonian in Eq.~(\ref{eq_hamiltonian})
into two parts,
\begin{eqnarray}
\label{eq_ham2}
H=H_0+H'(t),
\end{eqnarray}
where $H_0$ is time-independent and the time-dependent part is
contained in $H'(t)$.
In the non-interacting limit, $H_0$ is given by the first two terms in 
Eq.~(\ref{eq_hamiltonian}). In the weakly-interacting regime, 
$H_0$ has an additional contribution from the time-independent part of
the two-body coupling
\begin{eqnarray}
\label{eq_nipt}
V_{2b,0}(z_{j0})=\bar{g}\delta(z_j-z_0).
\end{eqnarray}
In both the $\bar{g}\rightarrow0$ limit and the
weakly-interacting regime, the time-dependent part of the Hamiltonian is
\begin{eqnarray}
H'(t)=\sum_{j=1}^{N}d\delta(z_j-z_0) \cos(\omega t) .
\end{eqnarray}

Following Ref.~\cite{zinner14, parish15, blume15}, we treat
the $\bar{g}\rightarrow\infty$ limit by replacing the $\delta$-function interaction
in Eq.~(\ref{eq_hamiltonian}) with 
a set of boundary conditions
enforced on the wave function $\Phi$ for (1,1) or (2,1) systems,
\begin{eqnarray}
  \label{eq_bpb}
\left(
\frac{\partial \Phi}{\partial z_{j0}}\bigg|_{z_{j0} \rightarrow 0^+} -
\frac{\partial \Phi}{\partial z_{j0}}\bigg|_{z_{j0} \rightarrow 0^-} 
\right )
= \frac{g m }{\hbar^2} \Phi|_{z_{j0} \rightarrow 0}.
\end{eqnarray}
$H'(t)$ can be written as
\begin{eqnarray}
\label{eq_hamt}
H'(t)=h \cos(\omega t) \mathcal{C},
\end{eqnarray}
where $\mathcal{C}$ is the contact density operator. The matrix elements of $\mathcal{C}$
can be obtained using the Hellmann-Feynman theorem with
Eq.~(\ref{eq_bpb})~\cite{zinner14, blume15}. They are given
in Eq.~(13) of Ref.~\cite{blume15}.

The (2,1) system is degenerate
in the limits $\bar{g}\rightarrow0$ and $\bar{g}\rightarrow\infty$. 
~\cite{zinner14, blume15}. 
In each degenerate manifold, we choose the states whose eigenvalues are smoothly connected
to the non-degenerate eigenvalues of the weakly perturbed system.
These ``good" eigenstates are obtained in the $\bar{g}\rightarrow0$ limit 
by applying a small perturbation 
given in Eq.~(\ref{eq_nipt}) and in the $\bar{g}\rightarrow\infty$ limit by 
applying the perturbative boundary condition given in Eq.~(\ref{eq_bpb})
to the degenerate manifolds.

We construct the Hermitian operator
\begin{eqnarray}
\mathcal{H}=H_0+H'(t)-i\hbar\frac{\partial}{\partial t}
\end{eqnarray}
and find the eigenvalues $\epsilon_n$ and the eigenfunctions
$|u_n(t)\rangle$ of the Hermitian operator $\mathcal{H}$ that satisfy
\begin{eqnarray}
\mathcal{H}|u_n(t)\rangle=\epsilon_n |u_n(t)\rangle.
\end{eqnarray}
The eigenfunctions $|u_n(t)\rangle$ are called Floquet modes and are
periodic in time
\begin{eqnarray}
|u_n(t+T)\rangle=|u_n(t)\rangle,
\end{eqnarray}
where $T=2\pi/\omega$ is the period.
The time evolution of an arbitrary state $|\Psi(t)\rangle$ can therefore be written
as
\begin{eqnarray}
|\Psi(t)\rangle=\sum_n c_n \exp(-i\epsilon_nt)|u_n(t)\rangle,
\end{eqnarray}
where the expansion coefficients $c_n$ are determined by the initial condition.

The Hilbert space of $\mathcal{H}$ can be expressed as the product space
$\mathcal{R}\otimes\mathcal{T}$, where $\mathcal{R}$ is the Hilbert space of $H_0$
and $\mathcal{T}$ is the space of functions of periodicity $T$.
A complete basis for $\mathcal{H}$ is formed from the product states
$|\psi_n\rangle\otimes|l\rangle$, where $|\psi_n\rangle$ are the eigenstates of $H_0$
and $\langle t|l\rangle=\exp(il\omega t)$.
Here, $l$ is an integer
quantum number that labels the ``Floquet band".
For systems considered in this work, the center-of-mass motion separates from
the relative motion and its time-dependence is trivial
because it is governed by a time-independent Hamiltonian. We hence focus on the
relative degrees of freedom and assume that the center-of-mass motion is 
always in its ground state.
For trapped (1,1) system, the exact solution for arbitrary interaction strength
is given in Ref.~\cite{busch98}. We use the exact solution combined with the time period
functions as our basis functions. For the trapped (2,1) system, the exact solutions
for the $\bar{g}\rightarrow0$ limit and the $\bar{g}\rightarrow\infty$ limit are known~\cite{zinner14}.
For the weakly-interacting regime, 
we use the exact solution for the $\bar{g}\rightarrow0$ limit combined
with the time period functions as our basis functions. 
In this work, we use about 100 basis functions for the time-independent Hamiltonian
and consider about 200 Floquet bands, corresponding to a total of 20,000 basis functions
for the time-dependent problem. We have tested to make sure that all our results remain 
unchanged if the basis set is further enlarged.

\section{Results}
\label{sec_result}

\subsection{$\bar{g}\rightarrow0$ limit}
\label{sec_result_ni}

\begin{figure}[htbp]
\includegraphics[angle=0,width=80mm]{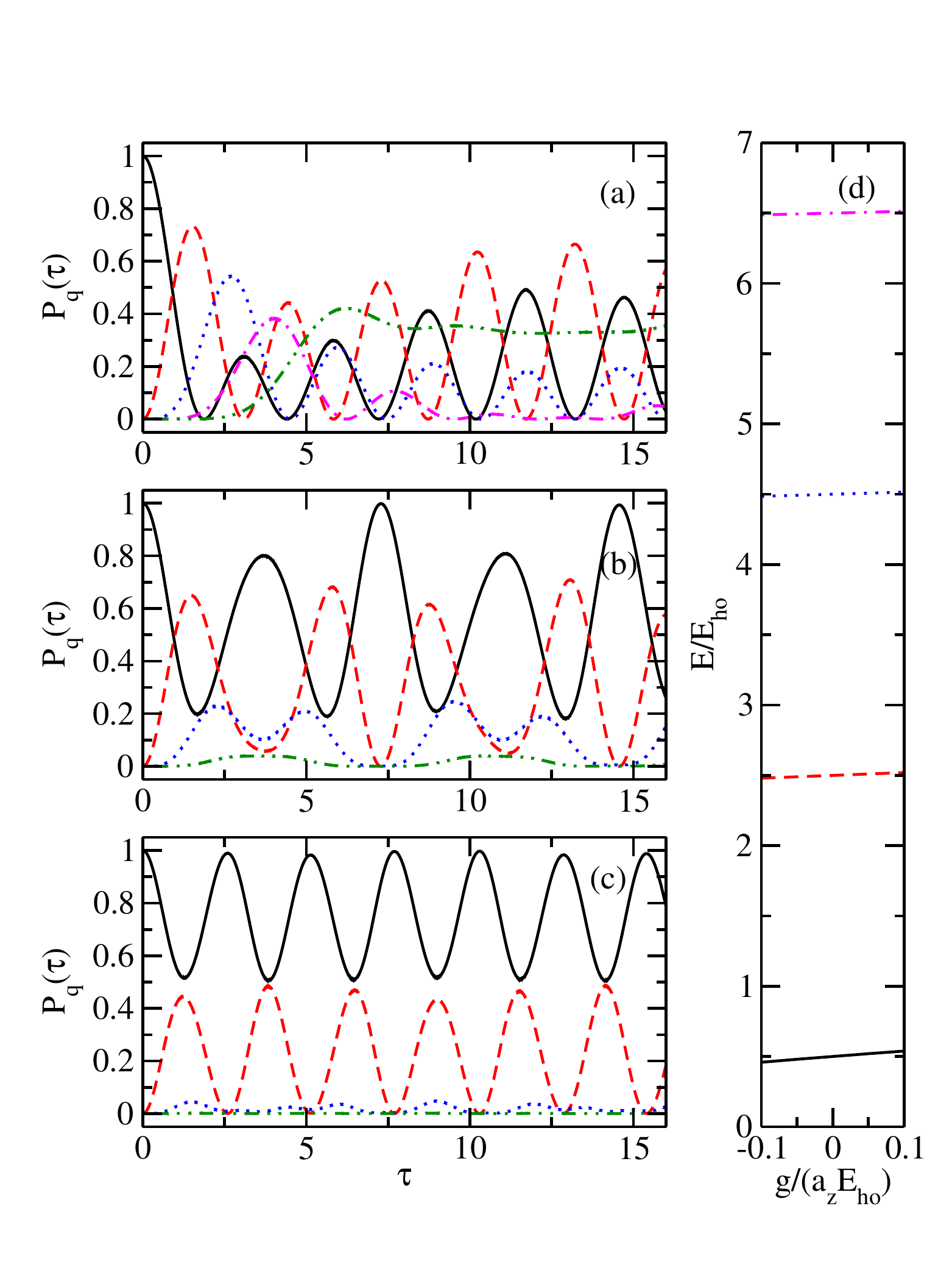}
\caption{(Color online) 
Dynamics of the (1,1) system under oscillating magnetic field in the $\bar{g}\rightarrow0$
limit. 
The driving strength is $d=0.01a_z E_\text{ho}$ for all cases. The figures
remain almost unchanged for different driving strength $d$ for 
$d<0.2a_zE_\text{ho}$.
Panel (a) shows the on-resonance case where $\omega=2\omega_z$.
The solid, dashed, dotted, dash-dotted 
and dash-dot-dotted lines show the
probability $P_q(\tau)$ of occupying the state with $q=0$, 1, 2, 3 and any other state
as functions of the scaled dimensionless time $\tau= t/(2\pi\hbar a_z/d)$.
Panel (b) and (c) show the slightly off-resonance cases, where the
detunings $\delta$ are $0.15d/(\hbar a_z)$ and $0.3d/(\hbar a_z)$, respectively.
The solid, dashed, dotted, 
and dash-dot-dotted lines show $P_q(\tau)$ for $q=0$, 1, 2, and any other state
as a function of $\tau$. 
Panel (d) shows the energy spectrum of (1,1) system near the $\bar{g}\rightarrow0$ limit
for even-parity states. The solid, dashed, dotted, and dash-dotted lines
show the energy of the four lowest states as a function of $g/(a_zE_\text{ho})$,
which are labeled by $q=0$, 1, 2, and 3.
}
\label{fig_ni11}
\end{figure}

In the $\bar{g}\rightarrow0$ limit, the relative eigenstates of the (1,1) system
for the time-independent part of the
Hamiltonian $H_0$ are the harmonic oscillator eigenstates
$\psi_q(z)=N_q\exp(-z^2/(4a_z^2))H_{2q}(z/(\sqrt{2}a_z))$,
with eigenenergies $E_q=(2q+1/2)\hbar\omega_z$.
Here, $z=z_1-z_0$ is the relative coordinate, $N_q$ is a
normalization constant, and $H_n$ denotes the $n$-th
Hermite polynomial.
The even-parity states are labeled by integers $q=0,1,2,\ldots$ and
are coupled by the time-dependent, zero-range interactions.
The energy spacing between two consecutive even-parity
eigenstates is $2\hbar\omega$. 
The odd-parity states, labeled by half integers
$q=1/2, 3/2, \ldots$, are not affected by the zero-range interaction
since the wave functions have zero amplitude at $z=0$.
Therefore, the odd-parity states will not be populated during the
time evolution.
A small portion of the energy spectrum for the (1,1) system near the
$\bar{g}\rightarrow0$ limit is shown in Figure~\ref{fig_ni11}(d).
Solid, dashed, dotted and dash-dotted lines show the eigenenergies of the even
parity states with $q=0, 1, 2$, and $3$, respectively, as a function of $g/(a_zE_\text{ho})$,
in the $\bar{g}\rightarrow0$ limit. even-parity states, whose eigenenergies 
remain constants for different $g$, are not shown in the figure. 

We prepare the system in the ground state $\psi_0(z)$ and 
time evolve the system under the Hamiltonian given in Eq.~(\ref{eq_ham2}).
The driving frequency $\omega$ is chosen to be on or near resonance
with the energy spacing between the ground state and the first excited even-parity
state $\psi_1(z)$.
We calculate the probability of occupying the even-parity states
$\psi_q(z)$ as a function of time. 
We find that
in the weak driving limit, which in practice translates to
$d/(a_z E_\text{ho})<0.2$,
the time dynamics is approximately invariant for different $d$
if we correspondingly scale the time by $2\pi\hbar a_z/d$.
This motivates us to define a scaled dimensionless time
$\tau= t/(2\pi\hbar a_z/d)$.
In this way, the dynamics we obtained in terms of $\tau$ applies to all
driving strengths $d$ in the range $d/(a_z E_\text{ho})<0.2$.
In the two-state model, this invariance is a consequence of the
validity of the first-order rotating wave approximation~\cite{rwa}. 
Our observation for the non-interacting (1,1) system is consistent with the
two-state case, and is also found to be true for other systems and
other regimes discussed in this work.

Figure~\ref{fig_ni11}(a) shows the probability of occupying different even-parity
states as a function of $\tau$
when the driving is on
resonance, i.e. $\omega=2\omega_z$. 
We find that not only is the $q=1$ state
populated, but higher-$q$ states are in turn populated with significant probability.
The probability of the populated states revives on the time scale set by
$2\pi\hbar a_z/d$, which is much longer than the driving period $T$.
The probability
of the ground state and the first excited states are, on average, the highest among 
all states and the
pattern of the revival dynamics roughly repeats itself in the long time limit. 
Figures~\ref{fig_ni11}(b) and (c) show the same quantity 
when the driving is slightly off-resonance, i.e. $\omega=2\omega_z+\delta$,
where $\delta$ is the detuning.
The dynamics is invariant for different choices of $d$
when the detuning $\delta$ is scaled by $d/(\hbar a_z)$
under the condition $\delta \ll \omega_z$. For example, the dynamics 
for the case shown in Fig.~\ref{fig_ni11}(c)
with $d=0.01a_zE_\text{ho}$ and $\delta=0.003\omega_z$
is almost the same as for $d=0.1a_zE_\text{ho}$ and $\delta=0.03\omega_z$.
The detuning suppresses the transition from the ground state to the excited states and
the suppression is stronger for higher energy states.
As the detuning increases, the system gradually becomes an effective two-state system
where the probability of occupying higher excited states becomes negligible.
The dynamics shown in Fig.~\ref{fig_ni11}(c) is very similar to the case where only
the $q=0$ and $1$ states are considered.

\begin{figure}[htbp]
\includegraphics[angle=0,width=80mm]{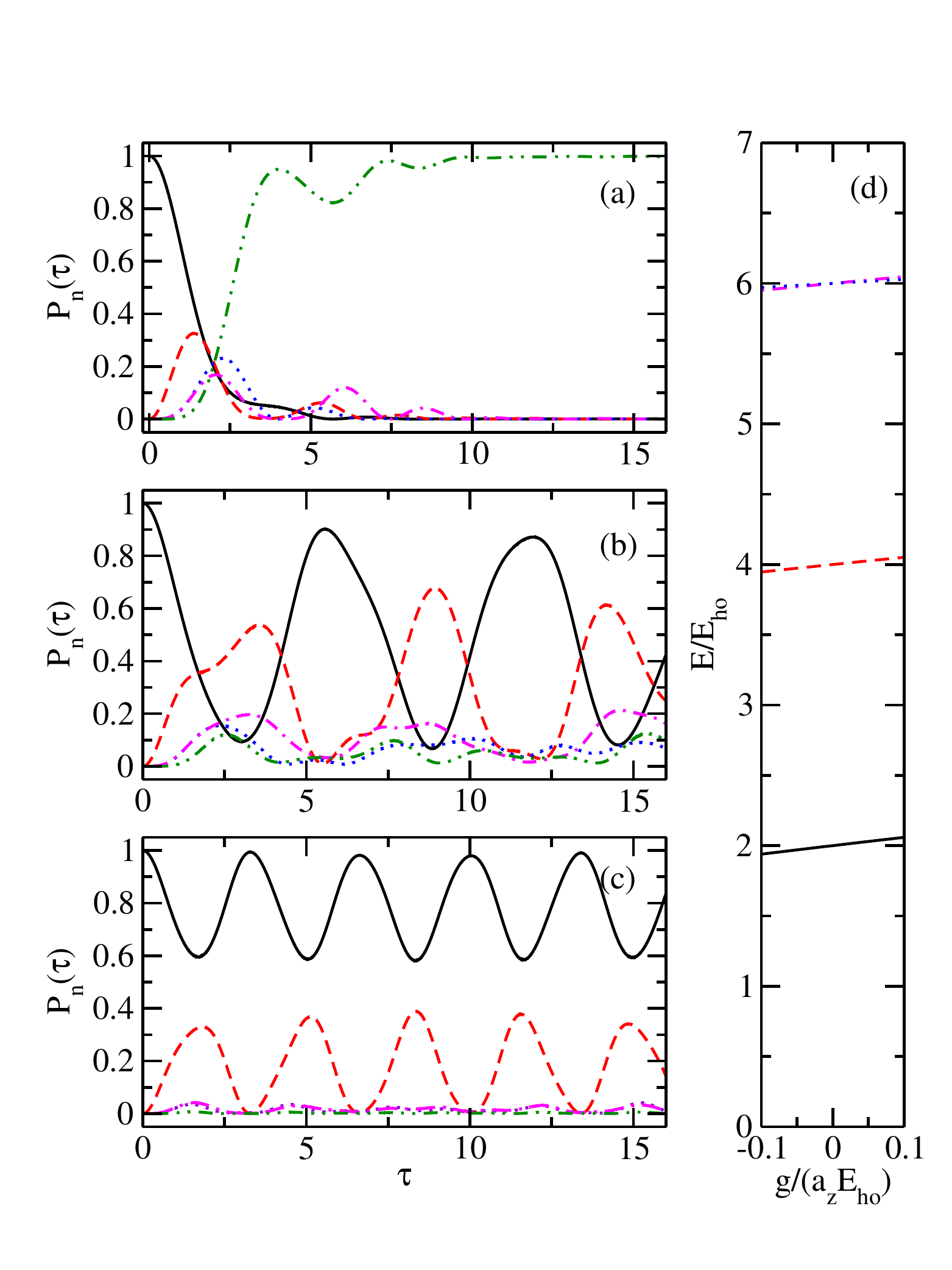}
\caption{(Color online) Dynamics of the (2,1) system under 
oscillating magnetic field in the $\bar{g}\rightarrow0$ limit. 
The driving strength is $d=0.01a_zE_\text{ho}$ for all cases. The figures
remains almost unchanged for different driving strength $d$ for 
$d<0.2a_zE_\text{ho}$.
Panel (a) shows the on-resonance case where $\omega=2\omega_z$.
The solid, dashed, dotted, dash-dotted 
and dash-dot-dotted lines show the
probability $P_n(t)$ of occupying the odd-parity states
with relative eigenenergies
$E_n/\hbar\omega_z=2$, $4$, $6$, $6$, and any other state
as functions of the scaled dimensionless time $\tau=t/(2\pi\hbar a_z/d)$. 
Note that there are two degenerate states with $E_n/\hbar\omega_z=6$
that are affected by the zero-range interaction
in the $\bar{g}\rightarrow0$ limit.
Panel (b) and (c) show the slightly off-resonance cases, where the
detunings $\delta$ are $0.15d/(\hbar a_z)$ and $0.3d/(\hbar a_z)$, respectively.
The solid, dashed, dotted, dash-dotted,
and dash-dot-dotted lines show the
probability $P_n(t)$ of occupying the
state with $E_n/\hbar\omega_z=2$, $4$, $6$, $6$, and any other state
as functions of $\tau$. 
Panel (d) shows the energy spectrum of the (2,1) system near $\bar{g}\rightarrow0$ limit
for states with odd relative parity. The solid, dashed, dotted, and dash-dotted lines
show the states with relative eigenenergies
$E_n/\hbar\omega_z=2$, $4$, $6$, and $6$.
Please note that the dotted and dash-dotted lines are very close to each other.
There are one and two additional states with $E_n/\hbar\omega_z=4$ and $6$, respectively,
that are not affected by the interaction and are not shown in the figure.
}
\label{fig_ni21}
\end{figure}

The ground state of (2,1) system has odd
relative parity.
Since the zero-range interactions conserve the relative parity,
the even-parity states will not be populated
if we initially prepare the state in the ground state. 
Hence, we consider only odd-parity states.
We note that, among the odd-parity states,
some are not affected by zero-range interaction.
Unlike the (1,1) system, degeneracy exists for the excited states
and is lifted in the presence of interaction.
The energy of the lowest relative eigenstate is $2\hbar\omega_z$.
There are two odd-parity states with energy $4\hbar\omega_z$.
Only one of these is affected by the zero-range interaction.
There are three odd-parity states with energy $6\hbar\omega_z$,
of which two states are affected by the zero-range interaction. 
A small portion of the energy spectrum for the (2,1) system near the
$\bar{g}\rightarrow0$ limit is shown in Fig.~\ref{fig_ni21}(d).
Solid, dashed, dotted and dash-dotted lines show the eigenenergies of odd-parity states 
with $E/E_\text{ho}=2, 4, 6$, and $6$, respectively, as a function of $g/(a_zE_\text{ho})$
in the $\bar{g}\rightarrow0$ limit. The states that are not affected by the zero-range
interactions are not shown. As explained in Sec.~\ref{sec_theory},
the two degenerate states with eigenenergy $6\hbar\omega_z$ in the $\bar{g}\rightarrow0$
limit are good eigenstates that change smoothly when deviating from the $\bar{g}\rightarrow0$
limit.

Figure~\ref{fig_ni21} shows the probability of occupying different odd-parity
states as a function of scaled dimensionless time $\tau$ for the (2,1) system.
When the driving is on
resonance, i.e.
$\omega=2\omega_z$, the dynamics is notably
different from the (1,1) system as the revival is much weaker
in the (2,1) system.
However, as the detuning increases, the transition to high excited states
is suppressed. As a result, the revival dynamics of the two lowest states
start to appear. Similar to the (1,1) system, 
higher detuning gradually leads to an effective two-state system as the transition
to higher excited states is further suppressed.

\subsection{Weakly-interacting regime}
\label{sec_result_wi}

As the system moves away from the
$\bar{g}\rightarrow0$ limit, the energy spacing between two consecutive
states of the (1,1) system is no longer $2\hbar\omega_z$.
The relative eigenstates with even-parity
for the time-independent Hamiltonian $H_0$
become $\psi_q(z)=N_q\exp(-z^2/(4a_z^2))U(-q,1/2,z^2/(2a_z^2))$
with eigenenergies $E_q=(2q+1/2)\hbar\omega_z$, for non-integer $q$. Here,
$N_q$ is a normalization constant and $U$ denotes the confluent 
hypergeometric function.
The non-integer quantum number $q$ is determined by the transcendental
equation~\cite{busch98}
\begin{eqnarray}
\frac{2\Gamma(-q+1/2)}{\Gamma(-q)}=-\frac{\bar{g}}{\sqrt{2}a_z \hbar \omega_z}.
\end{eqnarray}
The odd-parity states remain unaffected by the zero-range interaction.

\begin{figure}[htbp]
\includegraphics[angle=0,width=80mm]{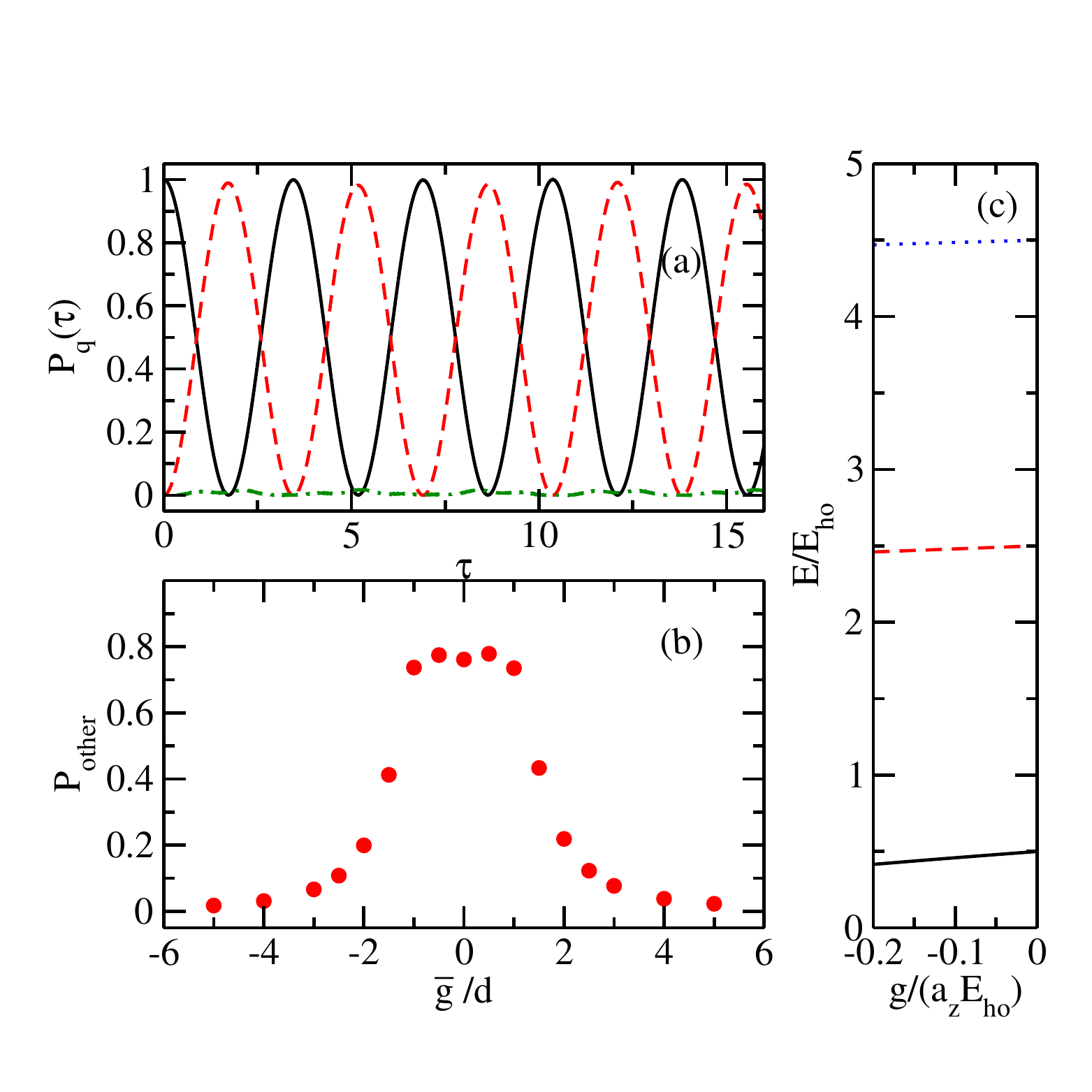}
\caption{(Color online) 
Dynamics of the (1,1) system under an
oscillating magnetic field in the weakly-interacting
regime.
Panel (a) shows the case where $\bar{g}=-0.1a_zE_\text{ho}$
and $d=0.02a_zE_\text{ho}$. The dynamics remains almost the
same for different values of $\bar{g}$ and $d$ giving the same 
$\bar{g}/d$.
Solid, dashed and dash-dotted lines show the 
probability $P_q(t)$ of occupying the ground state, the lowest
excited state with even parity, and any other state
as a function of the scaled dimensionless time $\tau=t/(2\pi\hbar a_z/d)$. 
Panel (b) shows the applicability of the two-state model.
Circles show the maximum probability of occupying any state other than
the two states in the Rabi oscillation $P_\text{other}$ as a function of $\bar{g}/d$.
Panel (c) shows the energy spectrum of the (1,1) system near 
$\bar{g}=-0.1a_zE_\text{ho}$ for even-parity states.
Solid, dashed, and dotted lines show the energy of 
the three lowest states with even parity.
}
\label{fig_wk11}
\end{figure}

As in the $\bar{g}\rightarrow0$ case, we prepare the system in the ground state.
We first consider an example case where
$\bar{g}=-0.1a_zE_\text{ho}$ and $d=0.02a_zE_\text{ho}$.
Since the interaction is weak and attractive, the energies of the relative eigenstates
are shifted down slightly compared to the $\bar{g}\rightarrow0$ case. 
The energies of the three lowest states with even-parity are $0.459\hbar\omega_z$,
$2.480\hbar\omega_z$, and $4.485\hbar\omega_z$, respectively,
giving spacings $2.021\hbar\omega_z$
and $2.005\hbar\omega_z$, respectively. 
The three lowest states are shown as solid, dashed, and dotted lines,
respectively, in Fig.~\ref{fig_wk11}(c).
We choose the driving frequency to be on resonance
with the energy difference between the ground state and the lowest excited state
with even parity, i.e. $\omega=2.021\omega_z$.
The difference in energy spacing between consecutive even-parity states greatly 
suppresses the
transition to higher excited states. As a result, the system is accurately described
by a two-state Rabi oscillation. 
Figure~\ref{fig_wk11}(a) shows the dynamics of the case discussed above.
The probability of occupying the states besides the two states in Rabi
oscillation is less than $2\%$. 
This observation is similar to
the three-dimensional system considered in Ref.~\cite{klaus06}.

Stronger interaction (greater $\bar{g}$) or weaker driving (smaller $d$)
suppress the transition to higher excited states,
resulting in a more ideal two-state model. With weaker interaction or 
stronger driving, the transition
to the higher excited states becomes non-negligible. 
We find that the dynamics of the weakly-interacting system remains almost invariant
for the same $\bar{g}/d$, e.g. the dynamics for a case with
$\bar{g}=-0.2a_zE_\text{ho}$ and $d=0.04a_zE_\text{ho}$ is almost identical
to the case shown in Fig.~\ref{fig_wk11}(a).
The applicability of the
two-state model can be assessed through the maximum probability of occupying any
state other than the two states in the Rabi oscillation $P_\text{other}$, which
are shown in Figure~\ref{fig_wk11}(b) for a range of average coupling
constant $\bar{g}/d$. 
The system can be described by a two-state model fairly well in the
$|\bar{g}|/d\gtrsim2.5$ regime, where $P_\text{other}$ is less than 10\%.
Besides the transition from the ground state to first excited state,
we checked that the two-state model is also applicable when driving a transition
from the ground state to other excited states.
The validity regime of the two-state model is greater for higher excited states.
For example, the transition from the ground state to the fourth excited state
and tenth excited state
can be described by a two-state model in the regime $|\bar{g}|/d\gtrsim1.0$
and $|\bar{g}|/d\gtrsim0.7$, respectively.

As shown above, the oscillating magnetic field can drive the system from 
the ground state to
an excited state with near-unity probability for a wide range of parameter
combinations. In cold atom experiments, the excited state could
be populated by the following scheme. 1. Prepare the system in
the $\bar{g}\rightarrow0$ limit ($\bar{g}\rightarrow0^{+}$). 2. Tune the interaction
strength to a small but non-zero positive value of $\bar{g}$ adiabatically through
the CIR. 3. Turn on the oscillating magnetic field, with the driving
frequency matching the energy difference between the ground state and
the excited state. The driving strength $d$ should be chosen according
to the guideline discussed above. Once the excited state is populated,
the system could be tuned further into a strongly-interacting regime
adiabatically through CIR.

\begin{figure}[htbp]
\includegraphics[angle=0,width=80mm]{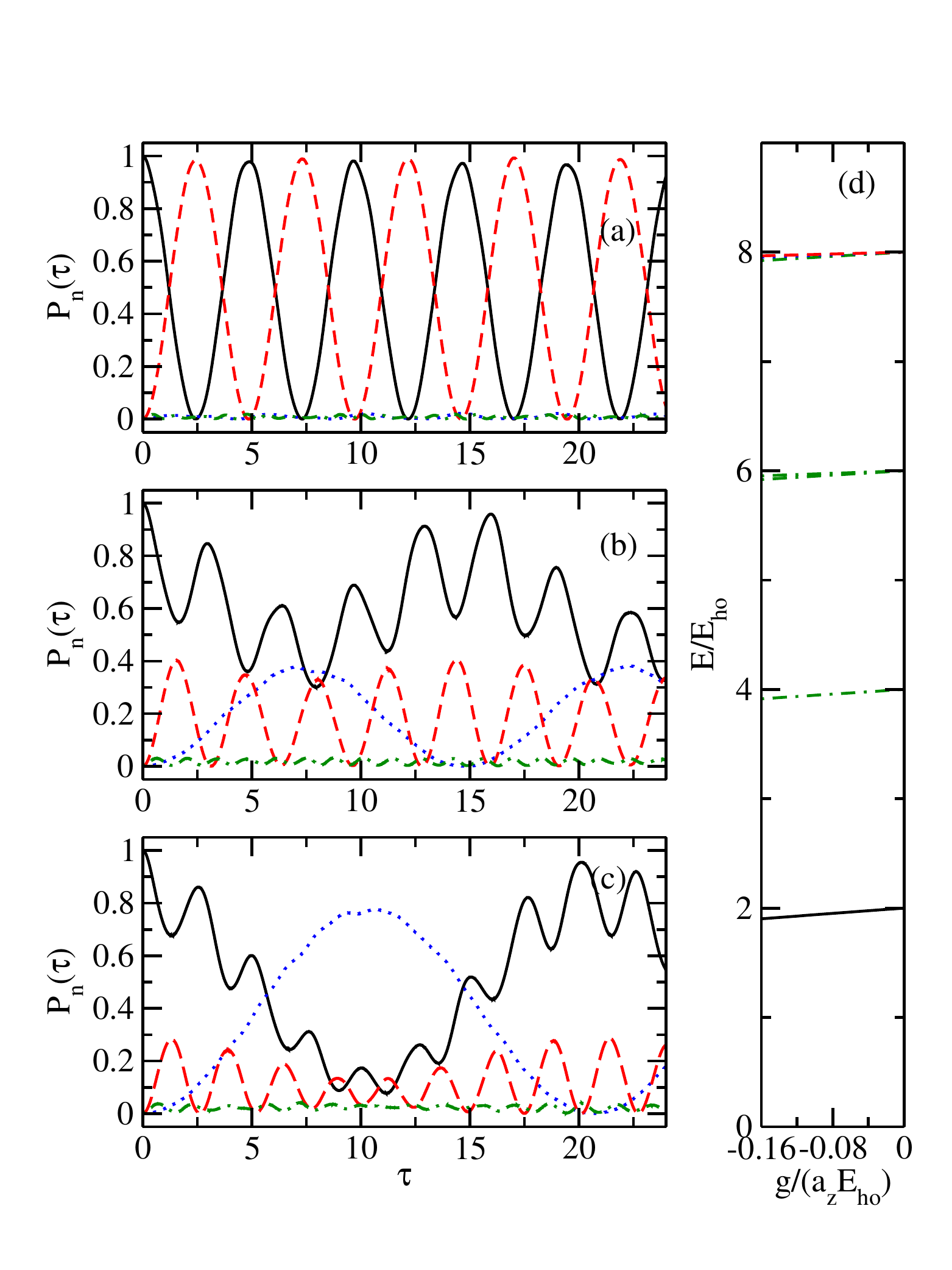}
\caption{(Color online) Dynamics of the (2,1) system under 
an oscillating magnetic field in the weakly-interacting
limit.  The average coupling 
constant is $\bar{g}=-0.08a_zE_\text{ho}$, and
the driving strength is $d=0.02a_zE_\text{ho}$.
Panel (a), (b), and (c) show cases with driving frequency
$\omega=6.0316\omega_z$, $6.0266\omega_z$ and $6.0251\omega_z$, respectively. Solid, dashed, dotted, and dash-dotted lines show the
probability $P_q(\tau)$
of occupying the ground state, $|\psi_b\rangle$, $|\psi_a\rangle$,
and any other state as
functions of the scaled dimensionless time $\tau= t/(2\pi\hbar a_z/d)$. See text for details.
Panel (d) shows the energy spectrum of the (2,1) system near $\bar{g}=-0.08a_zE_\text{ho}$
for states with odd parity. The solid, dashed, and dotted 
lines show the energy of ground state, $|\psi_b\rangle$, and $|\psi_a\rangle$, respectively.
The rest of the states are represented by dash-dotted lines. 
Please note that the dashed and dotted lines are very close to each other.
The states not affected by the
zero-range interactions are not shown in the figure.
}
\label{fig_wk21}
\end{figure}

For the interacting (2,1) system, the degeneracy of the excited states
in the $\bar{g}\rightarrow0$ limit is partly broken.
Some of the states within the degenerate manifold in the $\bar{g}\rightarrow0$ limit
become near-degenerate states in the weakly-interacting regime. 
For instance, there are four odd-parity
states with relative energy $8\hbar\omega_z$
in the $\bar{g}\rightarrow0$ limit. One of the four states is not affected by the zero-range
interaction while the other three are slightly shifted down by the weak attractive interaction.
For $\bar{g}=-0.08a_zE_\text{ho}$,
the energies of the three shifted states become $7.9620\hbar\omega_z$, $7.9764\hbar\omega_z$,
and $7.9833\hbar\omega_z$, respectively, giving energy spacings $0.0144\hbar\omega_z$
and $0.0069\hbar\omega_z$.
We note that the energy difference between the two
latter states, denoted here as $|\psi_a\rangle$ and $|\psi_b\rangle$,
is very small compared to $\hbar\omega_z$.
The on-resonance driving frequency from the ground state ($1.9516\hbar\omega_z$)
to $|\psi_a\rangle$ and $|\psi_b\rangle$
are  $6.0247\omega_z$ and $6.0316\omega_z$, respectively.
The ground state, $|\psi_b\rangle$, and $|\psi_a\rangle$ are shown
as solid, dashed, and dotted lines, respectively, in Fig.~\ref{fig_wk21}(d).
Other odd-parity states that are affected by the zero-range interactions are shown
as dash-dotted lines.
In this case, we will demonstrate that
it is possible to design a tunable effective three-state model by choosing a driving frequency 
that lies between the two on-resonance driving frequencies. 
We remark that the driving between the ground state and the two former states in the
near degenerate manifold does not yield an ideal three-state model as the case shown here since the
energy spacing is larger. 

Figure~\ref{fig_wk21} shows the dynamics of the (2,1) system in the parameter
combination discussed above.
Solid, dashed, dotted and dash-dotted lines show the 
probability of occupying the ground state,
 $|\psi_b\rangle$, $|\psi_a\rangle$, and all other odd-parity states as a function of time.
Panel (a) shows the case where the driving frequency $\omega$ is on resonance with the energy
difference between the ground state and $|\psi_b\rangle$, i.e. $\omega=6.0316\omega_z$.
In this case, the system can be described very well
by a two-state Rabi oscillation between the ground state
and $|\psi_b\rangle$. 
Panel (b) and (c) show the cases where the driving frequencies
are  $\omega=6.0266\omega_z$ and $6.0251\omega_z$, respectively.
In both cases, the ground state, $|\psi_a\rangle$ and $|\psi_b\rangle$ are populated with significant probability. The relative probability between 
$|\psi_a\rangle$ and $|\psi_b\rangle$ can be
tuned, within a certain range, by tuning the driving frequency.
The tunable effective three-state model could  potentially be used as a building block
for the realization of quantum Potts model in cold atom systems if
multiple trapped (2,1) systems are prepared and coupled together~\cite{potts}.

\textbf{\subsection{$\bar{g}\rightarrow\infty$ limit}
\label{sec_result_uni}}

\begin{figure}[htbp]
\includegraphics[angle=0,width=80mm]{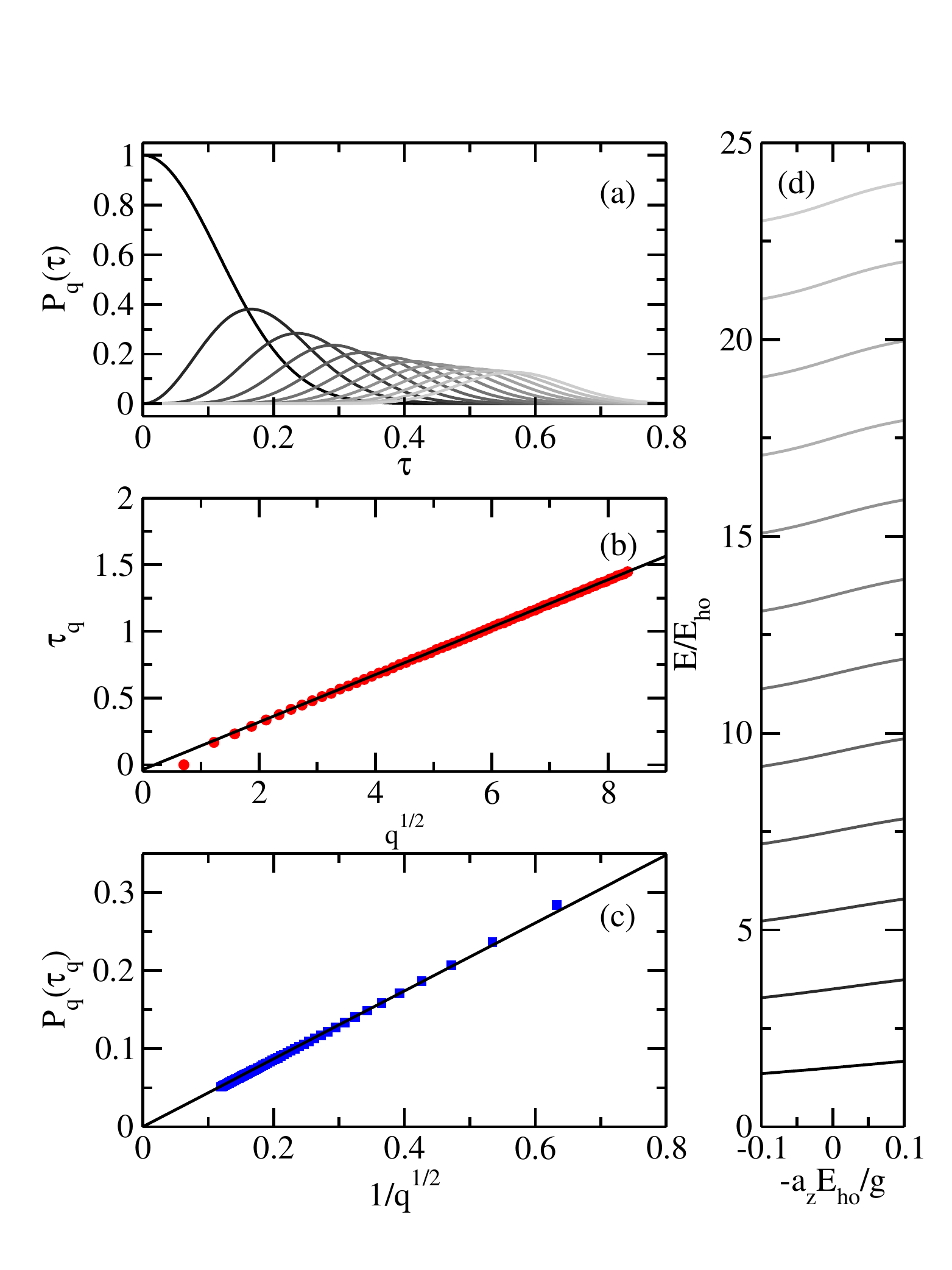}
\caption{(Color online) Dynamics of the (1,1) system under an oscillating magnetic field in the 
$\bar{g}\rightarrow\infty$ limit. The driving strength is $h=0.002/(a_zE_\text{ho})$.
The driving frequency is $\omega=2\omega_z$.
The solid lines in panel (a) show the probabilities $P_q(\tau)$ of occupying 12 lowest eigenstates
as functions of the scaled dimensionless time $\tau=t/(2\pi/(ha_z\hbar\omega_z^2))$. 
The eigenstates with quantum number $q=1/2$, $3/2$, $\ldots$ are shown 
successively by solid lines with
darker grey (peaking at a shorter time) to lighter grey (peaking at a longer time).
Circles in panel (b) show the scaled peaking time $\tau_q$ as a function of $\sqrt{q}$.
Squares in panel (c) show the peak probability $P_q(\tau_q)$ as a function of   
$1/\sqrt{q}$. The lines in panel (b) and (c) are linear fits.
Panel (d) shows the energy spectrum for the (1,1) system near the $\bar{g}\rightarrow\infty$ limit.
The states with lower (higher) energies are shown by solid lines with darker (lighter) grey.
}
\label{fig_uni11}
\end{figure}

In the $\bar{g}\rightarrow\infty$ limit,
the non-integer quantum number $q$ for the even-parity states of the (1,1) system
takes half integer values $q=1/2, 3/2, \ldots$ and the eigenenergies are given
by $E_q=(2q+1/2)\hbar\omega_z$. 
Hence, the even-parity states becomes degenerate with
the odd-parity states which are not affected by the zero-range interaction.
Figure~\ref{fig_uni11}(d) shows a small portion of the energy spectrum for the
(1,1) system near the $\bar{g}\rightarrow\infty$ limit. The odd-parity states are not shown in the figure.

We prepare the system in the lowest state with even parity
and let the system evolve under the Hamiltonian given in Eqs.~(\ref{eq_ham2}) and
(\ref{eq_hamt}).
We first set the driving frequency to be on resonance with the energy difference between
the two lowest states with even parity, i.e. $\omega=2\omega_z$.
We find that as the system evolves in time, every even-parity state in turn gets
excited and then depleted. 
Similar to the $\bar{g}\rightarrow0$ limit and the weakly-interacting regime,
the dynamics of the system is almost invariant for different driving strengths $h$
if we scale the time $t$ by $2\pi/(ha_z\hbar\omega_z^2)$.
Hence, in the $\bar{g}\rightarrow\infty$ limit, we define the scaled dimensionless time
$\tau=t/(2\pi/(ha_z\hbar\omega_z^2))$.
Solid lines in Figure~\ref{fig_uni11}(a) show the probability of 
occupying the 12 lowest eigenstates as a function of $\tau$. 
We find that the eigenstates with
larger $q$ quantum number peak at longer times $\tau_q$. 
Figure~\ref{fig_uni11}(b) shows the scaled peaking time $\tau_q$
of an eigenstate as a function of the square root of the quantum number   
$\sqrt{q}$. We find that $\tau_q$ scales linearly with $\sqrt{q}$. A two parameter
fit yields $\tau_q=-0.036(2)+0.178(1)\sqrt{q}$.
Figure~\ref{fig_uni11}(c) shows the peak probability $P_q(\tau_q)$ as a function of   
$1/\sqrt{q}$. 
We find that 
the peak probability is inversely proportional to $\sqrt{q}$. A two parameter fit
shows an absence of constant term within numerical accuracy and hence
$P_q(\tau_q)=0.435(1)/\sqrt{q}$.
We emphasize here that the two relations shown here are applicable in
the weak driving regime,
which translates to $h<0.05/(a_zE_\text{ho})$.
The two relations show that the energy is being continuously pumped into the system 
through the oscillating magnetic field and the coupling to higher excited states is unbounded.
The energy transfer between states in an open quantum system is a topic
of great interest~\cite{transport}. The system we considered is a specific
example that could be analyzed through the general framework in Ref.~\cite{transport}
using the quantum master equation.

\begin{figure}[htbp]
\includegraphics[angle=0,width=80mm]{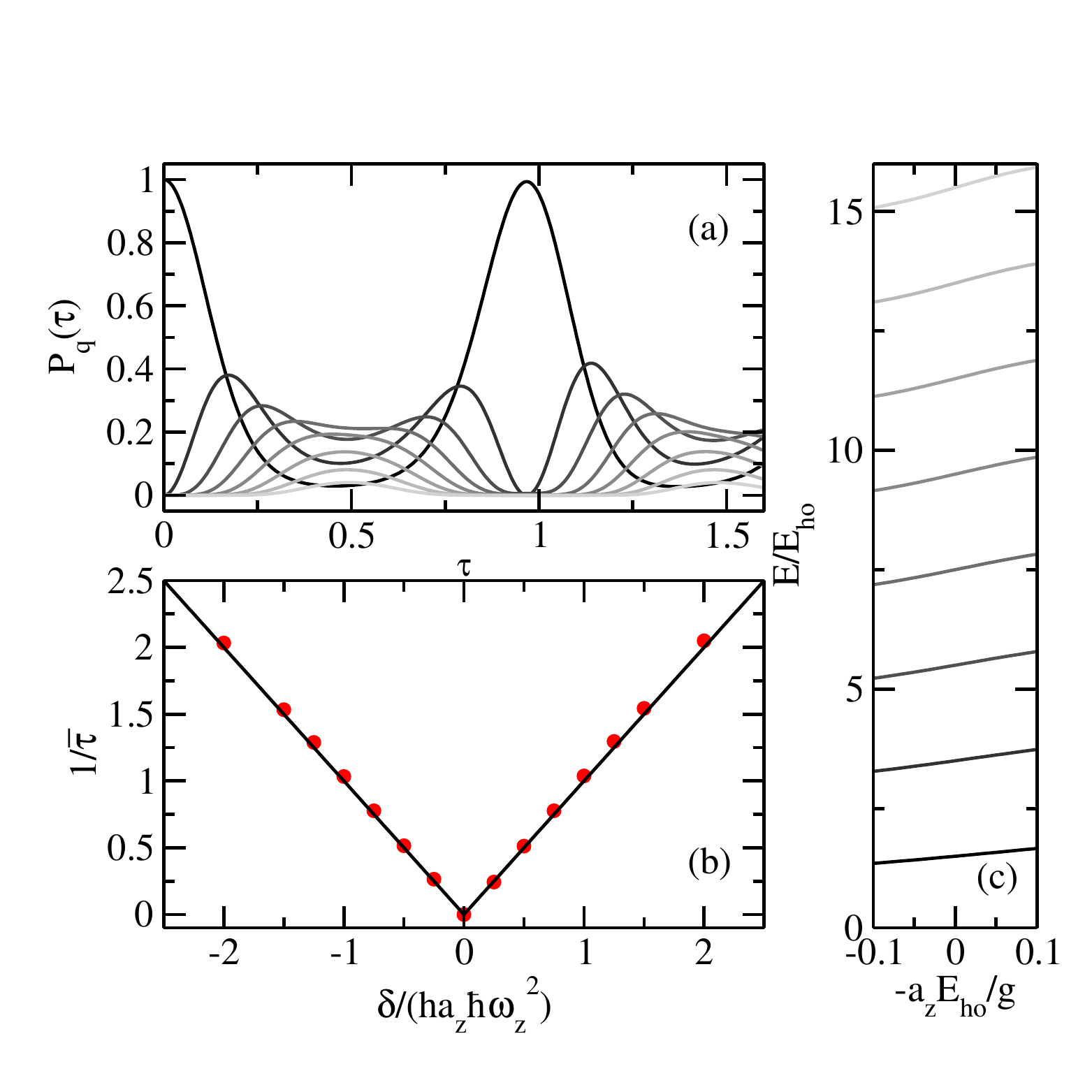}
\caption{(Color online) Dynamics of the (1,1) system under oscillating magnetic field in the 
$\bar{g}\rightarrow\infty$ limit. The driving strength is $h=0.002/(a_zE_\text{ho})$.
Panel (a) shows the case of near-resonant 
driving frequency $\omega=1.998\omega_z$. The detuning is
$\delta=1.0ha_z\hbar\omega_z^2$.
Solid lines show the probability of occupying the 8 lowest eigenstates
as a function of the scaled time $\tau=t/(2\pi/(ha_z\hbar\omega_z^2))$. 
The eigenstates with quantum number $q=1/2$, $3/2$, $\ldots$ are shown 
successively by solid lines with darker to lighter grey.
Panel (b) shows the inverse revival period $1/\bar{\tau}$ for the ground 
state as a function of scaled detuning $\delta/(ha_z\hbar\omega_z^2)$.
The solid line shows $1/\bar{\tau}=|\delta|/(ha_z\hbar\omega_z^2)$.
Panel (c) shows the energy spectrum for the (1,1) system near the $\bar{g}\rightarrow\infty$ limit.
The states with lower (higher) energies are shown by solid lines with darker (lighter) grey.
}
\label{fig_uni11s}
\end{figure}

The unbounded coupling is only present when the driving frequency is exactly on resonance.
When the driving is slightly off resonance, i.e.,
the detuning $\delta$ is non-zero, the revival of the ground state occurs on a time scale
determined by the detuning. Figure~\ref{fig_uni11s}(a) shows the dynamics of the
(1,1) system at $\bar{g}\rightarrow\infty$ under oscillating magnetic field with a driving frequency of 
$\omega=1.998\omega_z$, where the detuning is $\delta=1.0ha_z\hbar\omega_z^2$.
The number of excited states that are populated is limited compared to the
on-resonance case. 
The probability of the ground state revives after a certain time $\bar{\tau}$
to near unity. The revival dynamics repeats itself in a period of $\bar{\tau}$
in the long time limit. We calculated the revival period $\bar{\tau}$ for different
scaled detuning $\delta/(ha_z\hbar\omega_z^2)$.
Figure~\ref{fig_uni11s}(b) shows the inverse revival period $1/\bar{\tau}$
as a function of scaled detuning $\delta/(ha_z\hbar\omega_z^2)$.
We find that the revival period is inversely proportional to
the absolute value of the scaled detuning strength and the proportionality factor is
1 within numerical accuracy as determined from a fit, i.e, our numerics suggests that
$\bar{\tau}=ha_z\hbar\omega_z^2/|\delta|$. If we express this relation in
terms of unscaled revival period $T_\text{rev}$, we find $T_\text{rev}=2\pi/|\delta|$.

In the $\bar{g}\rightarrow\infty$ limit,
the dynamics of the (2,1) system is very similar to the (1,1) system. When the driving
is on resonance, i.e. $\omega=2\omega_z$, we observe similar unbounded coupling
as in the (1,1) system. 
Although the excited states of the (2,1) system at $\bar{g}\rightarrow\infty$ limit are degenerate, only one
good eigenstate
within each degenerate manifold can be populated with significant probability as 
time evolves. 
Similar to the (1,1) system, the scaled peaking time $\tau_n$ and 
the peak probability $P_n(\tau_n)$ are 
linearly proportional and reversely proportional to the square root
of eigenenergy $\sqrt{E_n}$, respectively.
Our fits yield $\tau_n=-0.10(2)+0.090(2)\sqrt{E_n/E_\text{ho}}$
and $P_n(\tau_n)=-0.06(1)+0.92(1)/\sqrt{E_n/E_\text{ho}}$.
The detuning has exactly the same effect on the dynamics as in the (1,1) system.
The relation between the revival period and detuning
discovered in the (1,1) system still holds for the (2,1) system.

\section{Conclusion and Outlook}
\label{sec_conclusion}
This paper studied the dynamics of a harmonically trapped two- and three-fermion systems
under an oscillating magnetic field. 
Previous works have studied the quantum dynamics of periodically driven one- and two-state
systems.
The systems considered in this work are considerably more complicated. Yet, they
represent the simplest realistic few-body systems
that have been successfully prepared in experiments. Therefore,
the dynamics studied in this work could be investigated using existing experimental setups. 
We analyzed these systems in the non-interacting
limit, weakly-interacting regime and the $\bar{g}\rightarrow\infty$ limit, and
we related these regimes
to the parameter regimes that are accessible in the experiments.

We found that the dynamics of the systems studied contains rich physics.
We revealed the dynamic difference in the the revival dynamics between the (1,1) and
(2,1) systems in the non-interacting limit.
We determined the parameter
regimes where the two-state Rabi oscillation model is applicable and
a population transfer with near-unity probability can potentially be realized
in the experiments.  
We also designed a tunable three-state model in the (2,1) system.
Moreover, we found an unbounded coupling to all excited states in the $\bar{g}\rightarrow\infty$ limit when
the driving is on resonance in both (1,1) and (2,1) systems. 
We discovered the relation between the peak probability,
peaking time, and the eigenenergy for all states.

Following the success in preparing few-body cold atom systems in the experiments, 
there has been increasing interest in the dynamics of
time-dependent few-body systems. 
Studies have been performed on the dynamics of a cold atom system after a rapid
change of scattering length~\cite{bohn16} and a displacement of the trap~\cite{alon14}.
Methods for state-transfer engineering have been proposed in cold atom systems~\cite{busch15}. 
There are also studies on dynamics for cold atom systems in a time-dependent
trapping potential~\cite{wu15, hammer16, blume16}.
Our study, which focuses on the dynamics induced by the time-dependent interactions,
belongs to the broader literature of few-body dynamics.

One of the key results in this work closely related to the experiments is that
population transfer between states with near-unity probability can be achieved
in the weakly-interacting regime. The reason is that the interaction breaks the
equal-energy spacing between eigenstates. Similar effects can be generated by
slightly deforming the trap, e.g. adding a term proportional to $z^4$ to the trapping
potential. At the same time, the time-dependent driving could also be introduced
by modulating confinement instead of the interaction. Although such methods may
introduce coupling between the center-of-mass motion and the relative motion,
the effect of coupling may be negligible in some parameter regimes.
These alternative techniques are worth investigating since they provide more flexibility
for the future experiments.
 
The additional energy scales introduced by the periodic driving open up
new possibilities in cold atom research. The few-body systems studied in
this work could potentially be coupled to form a lattice. The coupling of 
an array of effective two-state and three-state systems could lead
to new proposals in quantum computing. 
Recently, there has been a proposal to use cold atoms in shaking harmonic
traps to implement synthetic dimensions~\cite{synth} and
a realization of spin-orbit coupling through a time-modulated
magnetic field gradient~\cite{you15}. In those works, a single-particle 
picture for each harmonic trap is considered. Our work shows that
interactions between atoms can have an important impact on the dynamics. 
Our results for the few-body systems can serve as input for the coupled 
many-body systems in future studies.

\section{Acknowledgement}
X. Y. Y. is supported by
NSF Grant DMR-1309615, MURI Grant FP054294-D, and NASA Fundamental Physics Grant 1518233.
Y. Y. is supported by NSF Grant PHY-1415112.
X. Y. Y. thanks T.-L. Ho for useful discussions.
We are grateful to D. Blume for careful reading of the manuscript and thoughtful comments.

\bibliographystyle{apsrev4-1}
\nocite{apsrev41Control}
\bibliography{mybib}

\end{document}